\documentclass[useAMS,usenatbib]{mn2e}

\usepackage[dvipdfmx]{graphicx}

\title[Two types of LAEs]
{Two types of Lyman-alpha emitters envisaged from hierarchical galaxy formation}
\author[Shimizu and Umemura]{Ikkoh Shimizu,$^{1}$\thanks {E-mail:shimizu@ipmu.jp}
Masayuki Umemura$^{2}$ \thanks {E-mail:umemura@ccs.tsukuba.ac.jp} \\
$^{1}$Institute for the Physics and Mathematics of the Universe (IPMU), \\
The University of Tokyo, 5-1-5 Kashiwanoha, Kashiwa, Chiba 277-8582, Japan \\
$^{2}$Center for Computational Sciences, University of Tsukuba, Tsukuba 305-8577, Japan}
\begin{document}

\date{In original form 2009 July 1}

\pagerange{\pageref{firstpage}--\pageref{lastpage}} \pubyear{2009}

\maketitle

\label{firstpage}

\begin{abstract}
In the last decade, numerous Lyman-alpha emitters (LAEs) have been discovered 
with narrow-band filters at various redshifts. 
Recently, multi-wavelength observations of LAEs have been performed and 
revealed that while many LAEs appear to be young and less massive, 
a noticeable fraction of LAEs possess much older populations of stars 
and larger stellar mass. 
How these two classes of LAEs are concordant with the hierarchical galaxy formation 
scenario has not been understood clearly so far. 
In this paper, we model LAEs by three-dimensional cosmological simulations 
of dark halo merger in a $\Lambda$CDM universe. 
As a result, it is shown that the age of simulated LAEs can spread 
over a wide range from $2\times 10^6$yr to $9\times 10^8$yr. 
Also, we find that there are two types of LAEs, in one of which 
the young half-mass age is comparable to the mean age of stellar component, 
and in the other of which 
the young half-mass age is appreciably shorter than the mean age.
We define the former as Type 1 LAEs and the latter as Type 2 LAEs.
A Type 1 LAE corresponds to early starburst in a young galaxy, 
whereas a Type 2 LAE does to delayed starburst in an evolved galaxy,
as a consequence of delayed accretion of a subhalo onto a larger parent halo.
Thus, the same halo can experience a Type 2 LAE-phase as well as 
a Type 1 LAE-phase in the merger history. 
Type 1 LAEs are expected to be younger than $1.5 \times 10^8$yr, less dusty, 
and less massive with stellar mass $M_{\rm star} \la 5 \times 10^8 \rm ~M_{\odot}$,
while Type 2 LAEs are older than $1.5 \times 10^8$yr, even dustier,
and as massive as $M_{\rm star} \sim 5 \times 10^8 - 3\times 10^{10} \rm ~M_{\odot}$.
The fraction of Type 2s in all LAEs is a function of redshift,
which is less than 2 percent at $z \ga 4.5$,
$\sim$30 percent at redshift $z=3.1$, and 
$\sim$70 percent at $z=2$.
Type 2 LAEs can be discriminated clearly from Type 1s
in two color diagram of z'-H vs J-K.
We find that the brightness distribution of Ly$\alpha$ in Type 2 LAEs is 
more extended than the main stellar component, in contrast to
Type 1 LAEs.
This is not only because delayed starbursts tend to occur in the outskirts of
a parent galaxy, but also because Ly$\alpha$ photons are effectively absorbed 
by dust in an evolved galaxy. 
Hence, the extent of Ly$\alpha$ emission may be an additional measure 
to distinguish Type 2 LAEs from Type 1 LAEs. 
The sizes of Type 2 LAEs range from a few tens to a few hundreds kpc.
At lower redshifts, the number of more extended, older Type 2 LAEs increases.  
Furthermore, it is anticipated that the amplitude of angular correlation function 
for Type 2 LAEs is significantly higher than that for Type 1 LAEs, 
but comparable to that for Lyman break galaxies (LBGs). 
This implies that LBGs with strong Ly$\alpha$ line may include Type 2 LAEs.
\end{abstract}


\section{Introduction}
To explore the early evolutionary phases of galaxies is important 
to understand galaxy formation. 
\citet{PP1967} predicted that the starbursts in primeval galaxies emit 
significant Ly$\alpha$ emission 
through the recombination of ionized hydrogen in interstellar matter. 
Although many surveys attempted to discover such Ly$\alpha$ 
emitting galaxies (Ly$\alpha$ emitters: hereafter LAEs),  
but did not succeed to find them for a long time. 
In late 1990's, \citet{CH1998} discovered LAEs with narrow-band filters for the first time. 
Currently, numerous LAEs have been discovered at high redshifts $3 < z < 7$ 
by $8 \sim 10 {\rm ~m}$ class telescopes with narrow-band filters
\citep{Hu98, Hu99, Hu02,Kodaira03,Shimasaku03,Shimasaku06,Ha2004,Ou04,Ou05,Taniguchi05,Matsuda04,Matsuda05}.
Although the number of observed LAEs increases constantly, 
the nature of LAEs is still veiled.
Recently, surveys of LAEs in the various wavelength bands 
(optical, infrared, sub-mm, etc) have been performed actively 
\citep{Fin2007, Lai2008, Matsuda2007, Uchimoto2008, Fin2009, SMGLAE}, and
have revealed that while many LAEs appear to be young and less massive, 
a noticeable fraction of LAEs possess much older stellar populations and larger
stellar mass. 
We have not well understood how such two classes of LAEs are 
concordant with the hierarchical galaxy formation scenario. 
As for the physical origin of Ly$\alpha$ emission, 
the cooling radiation from a primordial collapsing cloud \citep{Haiman00,Fardal01}, 
from a galactic wind-driven shell \citep{TS00}, or from star-forming clouds 
in a young starburst galaxy \citep{MUF04} has been considered. 

Recently, \citet{MU2006} proposed a galaxy evolution scenario from LAEs to LBGs,
based on a supernova-dominated starburst galaxy model. In this scenario, LAEs
correspond to an early evolutionary phase of $<3 \times 10^8$yr.
Also, \citet{S2007} have constructed an analytic model of LAEs in a $\Lambda$CDM universe,
and found that if LAEs form in relatively low density regions of the universe
and the duration of starburst is as short as $0.7 \times 10^8$~yr,
the spatial distributions match the weak angular correlation function 
of LAEs observed at $z=3.1$. 
The spectral energy distribution (SED) fitting for observed LAEs has shown 
that LAEs mostly have young average age ($\sim 10^8$yr) and
low stellar mass ($10^8 \sim 10^9 \rm ~M_{\odot}$), 
and are less dusty or dust free \citep{Ga2006, Fin2007, Lai2008}. 
These young LAEs are consistent with the picture by \citet{MU2006}
and \citet{S2007}. 
Very recently, deep surveys of LAEs allow us to study detailed properties 
of individual LAEs. As a result, it has been revealed that 
LAEs have a wide range of age ($10^6 \sim 10^9$yr), 
stellar mass ($10^6 \sim 10^{10} {\rm ~M_{\odot}}$), and 
dust extinction with ${\rm A_{\rm V}}$ up to $1.3 \rm ~mag$ 
\citep{Fin2007, Lai2008, Fin2009}. 
LAEs detected in rest-frame optical/near infrared (NIR) bands tend to have older age, 
larger stellar mass, and stronger dust extinction than LAEs undetected in those bands.  
Thus, the picture of purely young starburst galaxies are not always 
reconciled with observed LAEs. So far, the physical reason has not 
been clarified for the existence of an old, massive, and dusty population of LAEs. 
The previous study has shown that 
a starburst-dominated galaxy can emit strong Ly$\alpha$ radiation in dust-free or
less dusty environments \citep{MU2006}. However, starburst galaxies cannot be
always LAEs in dusty environments (e.g. ultra-luminous infrared galaxies). 
Hence, what physical state corresponds to an old population of LAEs 
is an issue of great significance. 
Some authors argue that the clumpy distributions of dusty gas is 
important for the transfer of Ly$\alpha$ photons 
\citep{Neufeld1991, HansenOh2006, Fin2009c}. 
Since Ly$\alpha$ photons undergo resonant scatterings on the surface 
of gas clumps, photons can easily escape from the clumpy media.  
Such an effect provides the possibility of old, massive and dusty LAEs.  

In this paper, we explore how a young and old population of LAEs
are concordant with a hierarchical galaxy formation paradigm. 
For the purpose, we perform tree-dimensional cosmological simulations 
of dark halo merger in a $\Lambda$CDM universe,
incorporating the prescriptions of star formation, spectral evolution, and
dust extinction. 
Throughout this paper, 
we adopt $\Lambda$CDM cosmology with the matter density $\Omega_{\rm{M}} = 0.3$, 
the cosmological constant $\Omega_{\Lambda} = 0.7$, 
the Hubble constant $h = 0.7$ in units of $H_0 = 100 \rm{~km ~s^{-1} ~Mpc^{-1}}$, 
the baryon density $\Omega_{\rm B}h^2 = 0.02$, 
and $\sigma_8 = 0.92$ \citep{WMAP}.

\section{Model and Numerical Method}

\subsection{Basic Model}

To pursue the star formation history in the hierarchical galaxy formation,
we simulate the merging history of subgalactic halos 
(hereafter subhalo) by three-dimensional cosmological $N$-body simulations. 
Here, each particle is regarded as a subhalo that is 
supposed to consist of dark matter and baryons. 
We simulate $N = 256^3$ subhaloes in a comoving volume
of $(50 \rm ~Mpc)^3$. The mass of a subhalo is 
$2.73 \times 10^8 {\rm M_{\odot}}$. 
It is assumed that the star formation is triggered, 
when a subhalo accretes onto a parent halo. 
Then, we trace the stellar evolution separately for individual subhalos 
using a spectral synthesis code 'PEGASE' \citep{PEGASE}. 
Moreover, we take into account the effect of dust extinction on 
Ly$\alpha$ emission. 
The present approach allows us to analyze the distributions
of star forming regions in a halo, and also the clustering properties
of halos.

\begin{figure*}
\includegraphics[width = 110mm]{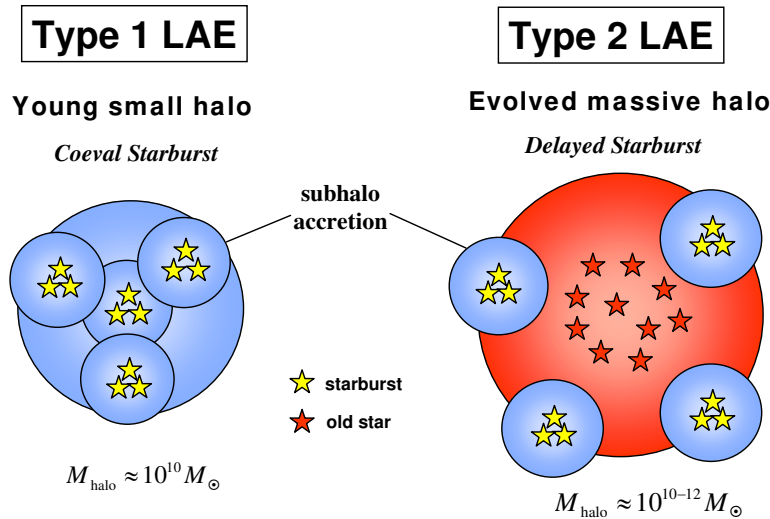}
\caption{Basic conception of a Type 1 and Type 2 LAE. 
A Type 1 LAE is an early phase of coeval starbursts in a young small halo. 
A Type 2 LAE corresponds to delayed starbursts in an evolved massive halo.}
\label{LAEs_Type}
\end{figure*}

There are basically two types of subhalo accretion.
One is the almost contemporaneous accretion of subhaloes in
a young small parent halo, 
and then coeval starbursts take place in the halo. 
The other is the delayed accretion onto an evolved massive halo,
and then a newly-triggered starburst and an old stellar population coexist.
Both types have the potentiality of becoming LAEs. 
If they satisfy LAE conditions (see the detail below), 
we call the former Type 1 LAEs and the latter Type 2 LAEs.
A schematic view of Type 1 and Type 2 LAEs is shown in
Fig. \ref{LAEs_Type}. 

\subsection{Numerical Method}

We perform a cosmological $N$-body simulation 
with the particle-particle-particle-mesh ($\rm P^3M$) algorithm \citep{P3M}. 
The numerical scheme is based on \citet{Y2000}. 
The size of comoving simulation box ($L_{\rm box}$) 
is set to be the same as the size of LAE survey at $z=3.1$ by \citet{Ha2004}, 
that is, $50 \rm ~Mpc$ in linear scale. 
This allows us to adjust LAE conditions by directly comparing
with the observation.
Here, the periodic boundary condition is imposed. 
We use the Plummer softening function for gravitational force, 
with the softening length of  $\epsilon_{\rm grav}=L_{\rm box} / (10N^{1/3})$ 
($\sim 20 {\rm ~kpc}$ in a comoving scale). 

A parent halo is found using 
a friends-of-friends algorithm \citep{FOF} 
with linking length equal to 0.2 of the mean particle separation. 
In this study, a system with $\geq 2.7 \times 10^{10} \rm M_{\odot}$ 
(say, equal to or more than 100 particles) is identified as a parent halo 
so that the system mass would be corresponding to observed LAEs. 
As previously mentioned, 
each particle (subhalo) in a parent halo has individual age. 
It is assumed that the star formation is triggered, when a subhalo 
accretes onto a parent halo. 
The star formation occurs only in the subhalo, and therefore
no star formation is triggered at the central of host halo.
Also, a subhalo which underwent the star formation once
does not trigger the star formation again.
Here, the star formation in each subhalo is assumed to occur at the rate as
\begin{equation}
\psi(t) = f_{\rm eff} \exp{\left(-\frac{t}{\tau_{\rm s}} \right)}.
\end{equation}
The star formation timescale $\tau_{\rm s}$ is set to match
the typical lifetime of young LAEs predicted by \citet{MU2006} and \citet{S2007},
that is, $1.0 \times 10^8 \rm yr$.
The efficiency $f_{\rm eff}$ is determined so that the final fraction of stellar mass
to total baryonic mass is 10 percent. Here, the Salpeter initial mass function
is assumed. In the present model, each particle is supposed to consist of dark 
matter and baryons. As far as the model LAEs concerned, newly accreted
subhalos are predominantly responsible for Lyman alpha emission, 
and those subhalos emits the bulk of UV photons before their dynamical
relaxation. Nonetheless, two-component (dark matter and baryon)
simulations are very worth investigating in the future work.

We identify LAEs under the same conditions as \citet{Ha2004}, that is,
$L_{\rm Ly\alpha, obs} \geq 1.4 \times 10^{42} {\rm erg / s}$
and $EW_{\rm rest} \geq 20 {\rm  \AA}$,
where $L_{\rm Ly\alpha}$ and $EW_{\rm rest}$ are observed Ly$\alpha$ luminosity and 
Ly$\alpha$ equivalent width at rest-frame, respectively. 
We estimate observed Ly$\alpha$ luminosity by 
\begin{equation}
L_{\rm Ly\alpha, obs} = f_{\rm esc} L_{\rm Ly\alpha, int}, 
\end{equation}
where $L_{\rm Ly\alpha, int}$ is intrinsic Ly$\alpha$ luminosity,
and $f_{\rm esc}$ is the escape fraction of Ly$\alpha$ photons.
We calculate $L_{\rm Ly\alpha, int}$ using "PEGASE" \citep{PEGASE}. 
The escape fraction is evaluated in terms of dust extinction as
\begin{equation}
f_{\rm esc} = \exp{(-\tau_{\rm dust})}, 
\end{equation}
where $\tau_{\rm dust}$ is 
the line-of-sight optical depth of dust with opacity 
in proportion to local metallicity. 
In practice, $\tau_{\rm dust}$ is dependent on grain size distributions,
which are not well known in high-$z$ objects. Hence, we normalize the net
cross section of dust grains so that the number of simulated LAEs should 
match the observed number of LAEs at $z=3.1$ \citep{Ha2004}. 
(In the present analysis, dust is treated as a pure absorber of Ly$\alpha$ photons,
and the scattering of Ly$\alpha$ photons is not solved. 
The treatment of dust extinction can be more
sophisticated in the future analysis.)

\section{Results}
\subsection{Two Types of Simulated LAEs}
\begin{figure}
\includegraphics[width = 80mm]{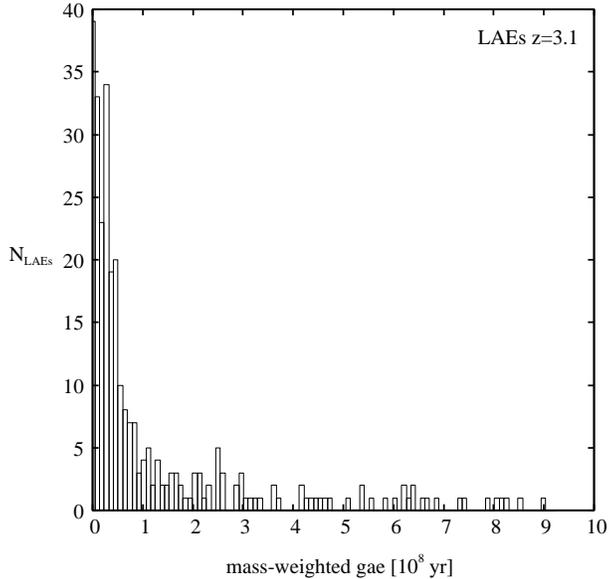}
\caption{Age distributions of simulated LAEs. 
The age is defined by the mass-weighted age.}
\label{Age}
\end{figure}

\begin{figure}
\includegraphics[width = 70mm]{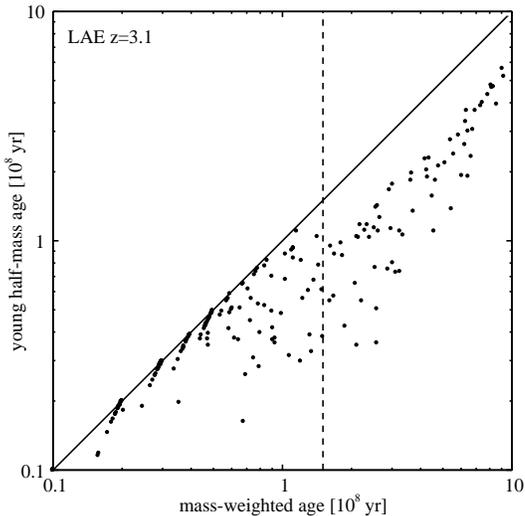}
\caption{Young half-mass ages against mass-weighted mean ages.
The straight line denotes the equality of two age definitions.
A vertical dashed line shows $1.5 \times 10^8$yr. 
We define LAEs younger than $1.5 \times 10^8$yr as Type 1s, and
older ones as Type 2s. }
\label{def}
\end{figure}

\begin{figure}
\includegraphics[width = 80mm]{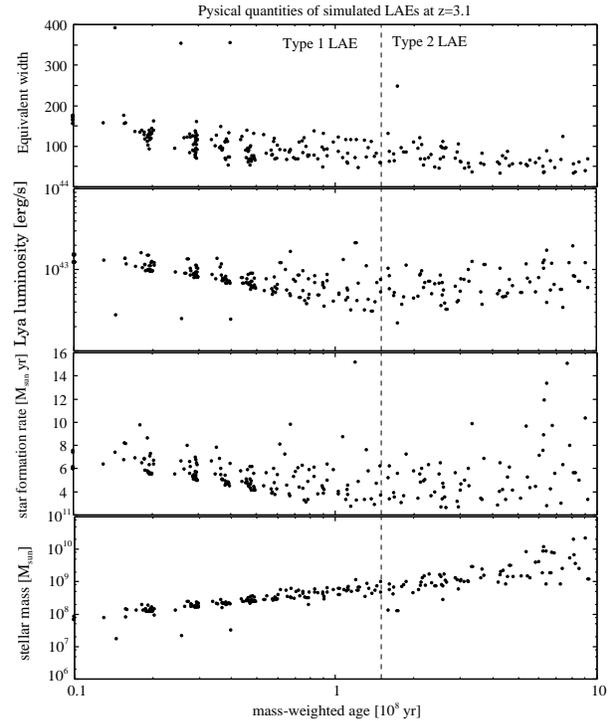}
\caption{Equivalent width, Ly$\alpha$ luminosity, star formation rate, 
and stellar mass against the mass-weighted age for simulated LAEs.
A vertical dashed line shows $1.5 \times 10^8$yr. }
\label{properties}
\end{figure}

\begin{figure}
\includegraphics[width = 70mm]{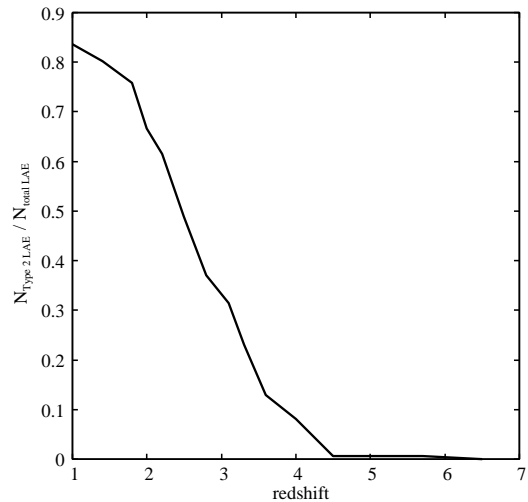}
\caption{The fraction of Type 2 LAEs at $1 < z < 7$. The fraction is defined 
by $N_{\rm Type~ 2~ LAEs} / N_{\rm total~ LAEs}$, where
$N_{\rm Type~ 2~ LAEs}$ and $N_{\rm total~ LAEs}$ are the number of 
Type 2 LAEs and all simulated LAEs, respectively.}
\label{frac}
\end{figure}

In Fig. \ref{Age}, the distributions of mass-weighted age are shown
for simulated LAEs. 
The ages spread widely from $2\times 10^6$yr to $9\times 10^8$yr.
Interestingly, a significant number of LAEs are much older than
$\approx 10^8$yr that is the lifetime of young LAEs predicted 
by \citet{MU2006} and \citet{S2007}.
As seen in Fig. \ref{Age}, the distributions are fairly continuous 
from younger LAEs to older ones. 
Younger LAEs are early coeval starburst galaxies, while
older LAEs result from delayed starbursts 
triggered by later subhalo accretion onto evolved halos. 
In order to discriminate delayed starbursts from coeval
young starbursts, we plot the young half-mass ages against 
the mass-weighted mean ages in Fig. \ref{def},
where, the young half-mass age is defined as 
the mass-weighted age of the young half subhalos included in a host halo.
If starbursts are coeval in a halo, young half-mass ages 
should be comparable to mean ages. 
It is clearly seen in Fig. \ref{def}
that in the part older than $\sim 1.5 \times 10^8$yr, 
coeval starbursts disappear and only delayed starbursts appear to
be taking place. Hence, in this paper, we define
LAEs younger than $1.5 \times 10^8$yr as Type 1 LAEs, and
older ones as Type 2 LAEs. 
It is noted that there is not a clear boundary of two classes
at $1.5 \times 10^8$yr, 
but the transition is actually gradual in the sense that
coeval and delayed starbursts are blended 
around $\approx 10^8$yr. 
Nevertheless, as shown below, we find distinctive trends
in photometric properties between Type 1s and Type 2s 
defined here. 

Fig. \ref{properties} shows 
equivalent width of Ly$\alpha$ emission, Ly$\alpha$ luminosity, 
star formation rate, and stellar mass against the mass-weighted age 
for simulated LAEs. 
Equivalent width (EW) decreases with ages for Type 1 LAEs, ranging 
from 40\AA\, to 200\AA,
although some are at a level of 350-400\AA. 
EW for Type 2s is randomly distributed in the range of 30\AA-150\AA.
Ly$\alpha$ luminosity is basically in proportion to
star formation rate for Type 1 LAEs, and gradually decreases with ages.
For Type 2 LAEs, Ly$\alpha$ luminosity randomly spread
in the range of $\sim 2 \times 10^{42}{\rm erg~s^{-1}}$ 
to $\sim 2 \times 10^{43}{\rm erg~s^{-1}}$.
Interestingly, in old Type 2 LAEs ($>6 \times 10^8$yr),
a high star formation rate does not always lead to high Ly$\alpha$ luminosity.
This can be understood by the effect of dust extinction as argued below.

The stellar mass $M_{\rm star}$ of LAEs is a fairly monotonic function of age.
Type 1 LAEs are less massive with $M_{\rm star} \la 5 \times 10^8 \rm ~M_{\odot}$,
while Type 2s are as massive as 
$M_{\rm star} \sim 5 \times 10^8 - 3\times 10^{10} \rm ~M_{\odot}$.
Recent observations show that
LAEs detected by rest-frame optical/NIR bands are more massive 
than $10^9~ \rm M_{\odot}$ and pretty older
\citep{Fin2007, Lai2008, Matsuda2007, Uchimoto2008, Fin2009}.
Such objects may correspond to Type 2 LAEs.

The fraction of Type 2 LAEs in all LAEs is predicted to be a function of redshift. 
Fig. \ref{frac} represents the Type 2 LAE fraction against redshift.  
Obviously, the Type 2 fraction increases with decreasing redshift.
At $z \ga 4.5$, the Type 2 fraction is less than 2 percent, since massive haloes 
have not grown yet, whereas it is
$\sim$30 percent at redshift $z=3.1$ and 
$\sim$70 percent at $z=2$.
This trend is concordant with the recent observations by \citet{Nilsson2009}.

\subsection{Spectrophotometric Properties of Simulated LAEs}

\begin{figure}
\includegraphics[width = 70mm]{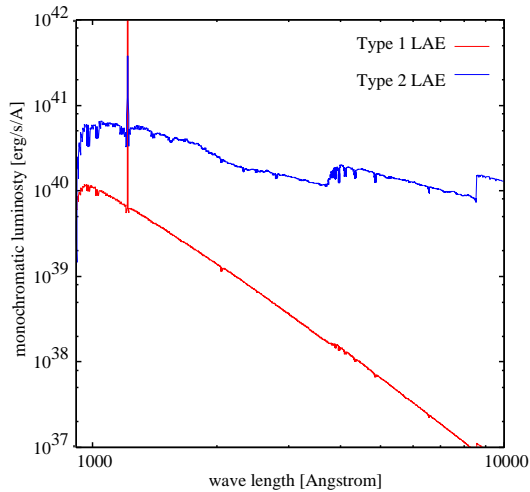}
\caption{Spectral energy distributions of a typical Type 1 and Type 2 LAE.
A red line represents Type 1 and a blue line does Type 2.}
\label{SED}
\end{figure}

\begin{figure*}
\includegraphics[width = 70mm]{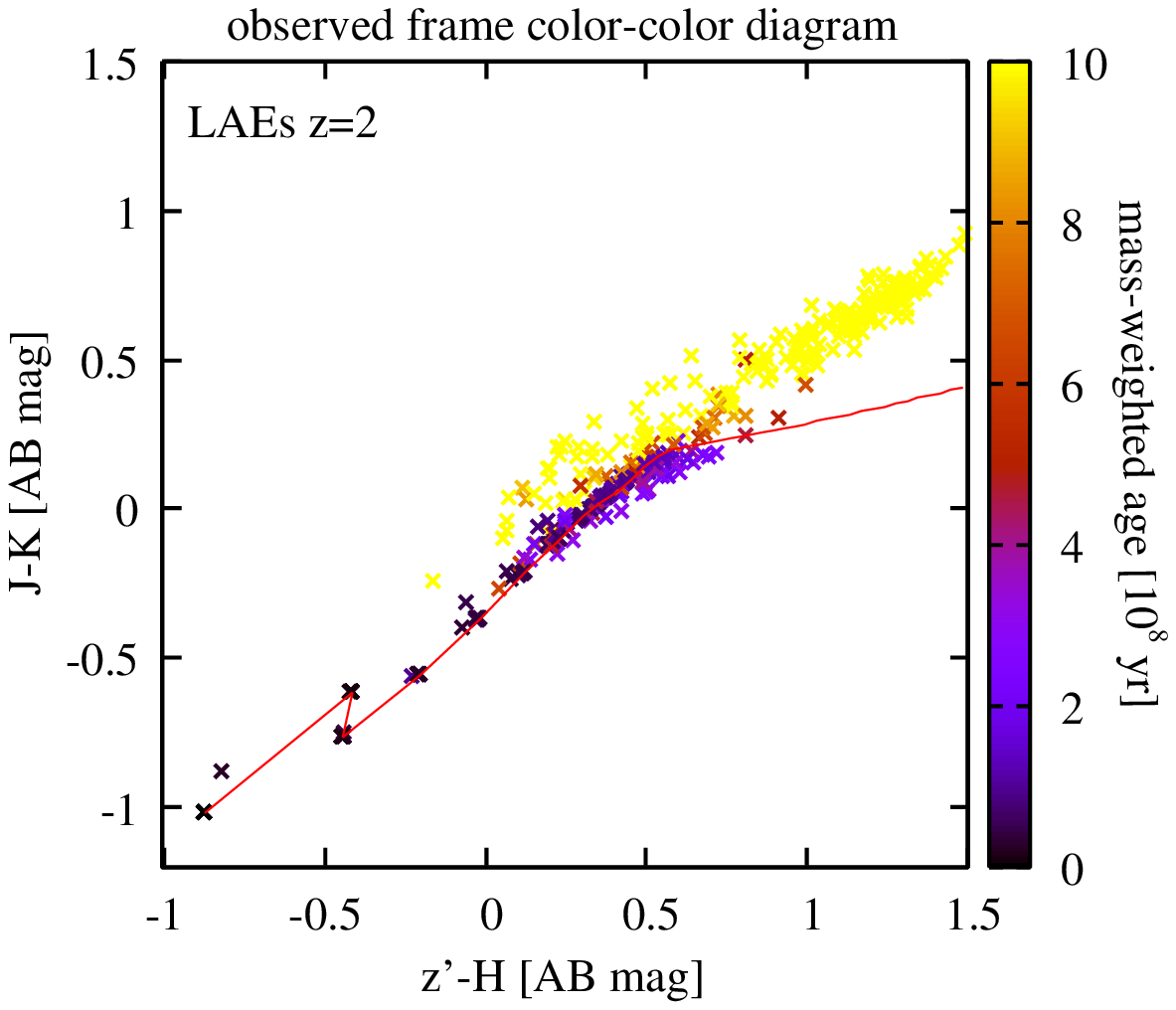}
\includegraphics[width = 70mm]{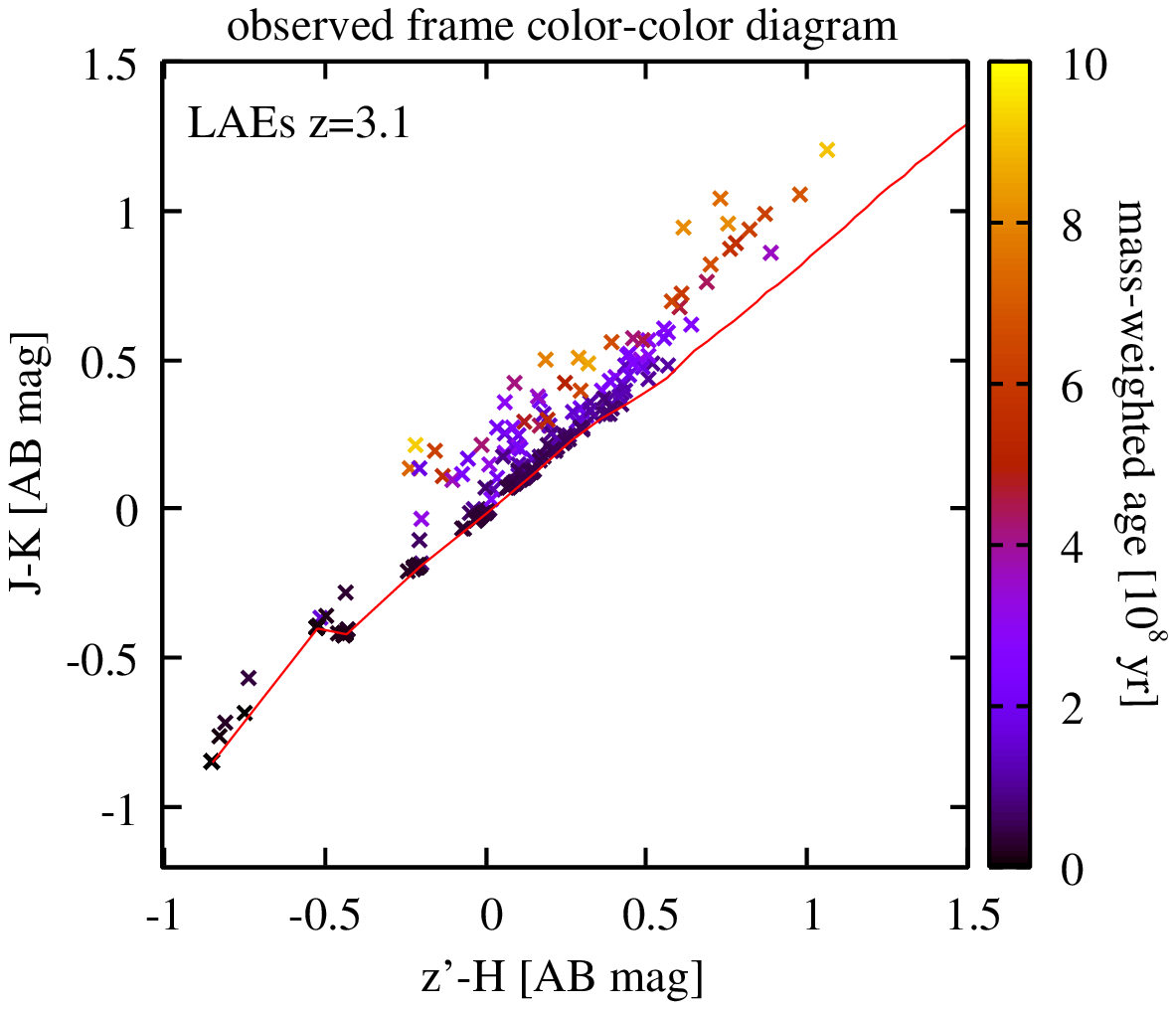}
\includegraphics[width = 70mm]{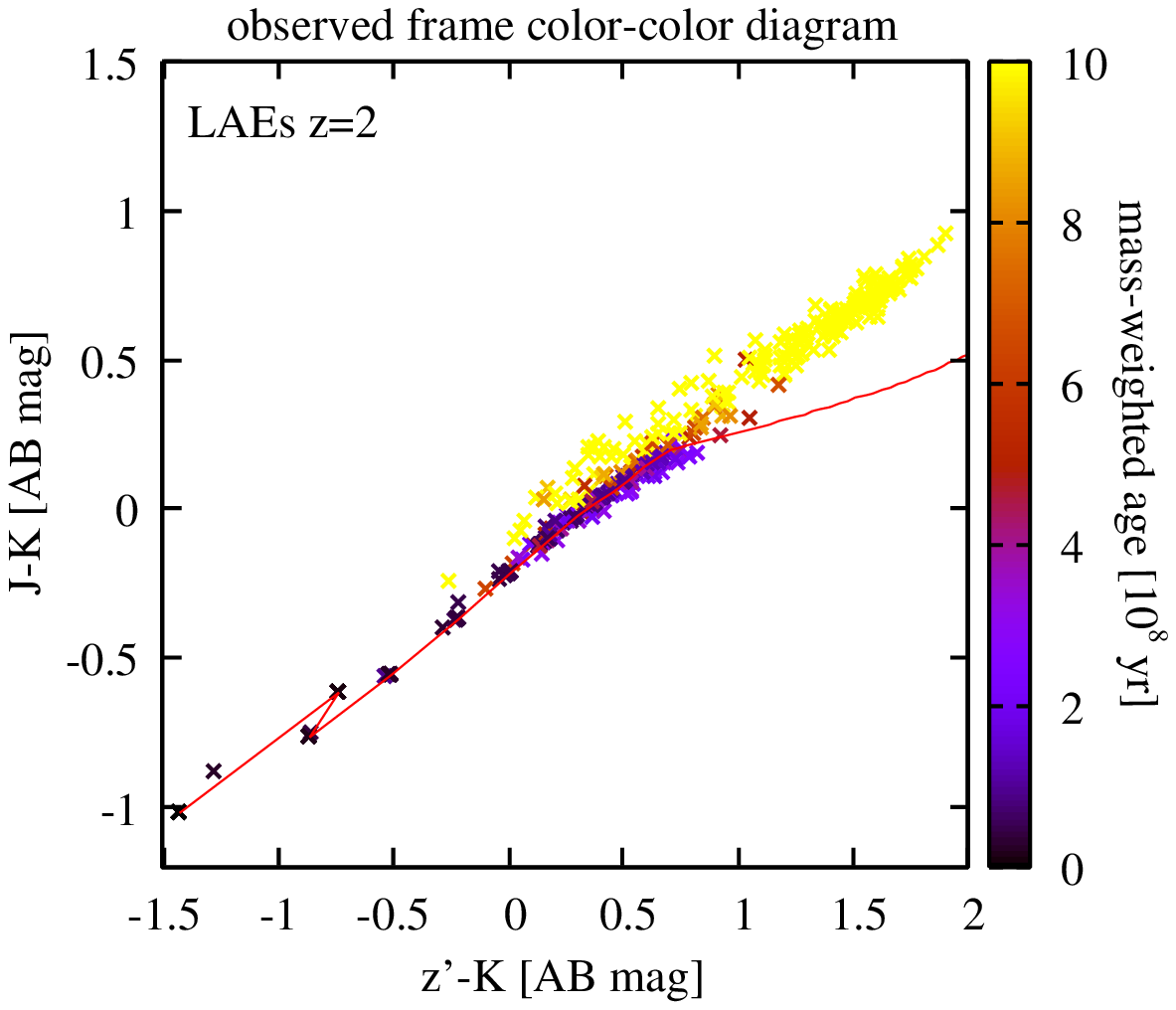}
\includegraphics[width = 70mm]{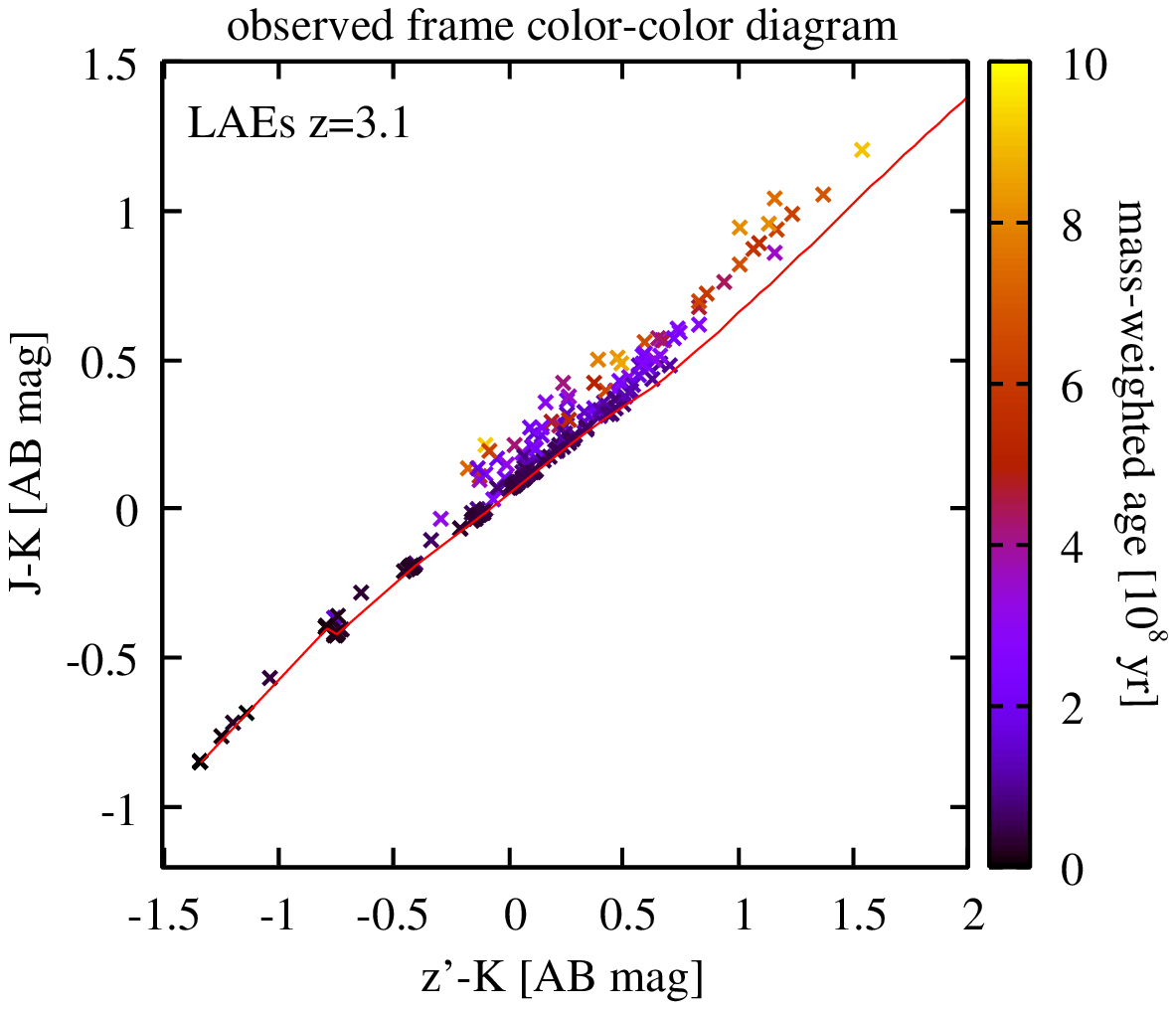}
\caption{Two color diagrams in observed frames. 
The upper two panels are z'-H vs J-K,
where the left panel is shown for LAEs at $z=2$ and the right panel is at $z=3.1$.
The lower two panels are z'-K vs J-K,
where the left panel is shown for LAEs at $z=2$ and the right panel is at $z=3.1$.
A red line in each panel denotes the evolutionary track of a single
starburst. The mass-weighted ages are shown by colors with
an attached color legend bar.}
\label{twocolor}
\end{figure*}

In Fig. \ref{SED}, we show
spectral energy distributions (SEDs) of a typical Type 1 and Type 2 LAE,
which are calculated by "PEGASE" \citep{PEGASE}. 
A Type 1 LAE is composed of young stars, while the host galaxy of
a Type 2 LAE is dominated by evolved stars, where a distinct 4000 \AA\, break
is seen. Accordingly, the colors of two types of LAEs are different.

In Fig. \ref{twocolor}, two color diagrams are shown in observed frames 
for $z=2$ LAEs and $z=3.1$ LAEs.
The upper two panels are z'-H vs J-K, while 
the lower two panels are z'-K vs J-K,
where z'-band is $\approx$9000\AA, J-band $\approx$12000\AA,
H-band $\approx$17000\AA, and K-band $\approx$22000\AA.
A red line in each panel indicates the evolutionary track of stars born
in a single starburst. Therefore, LAEs near this line represent
almost coeval starbursts. The mass-weighted ages are shown by colors.
As expected, Type 1 LAEs ($\leq 1.5 \times 10^8$yr) 
well follow the single starburst track,
whereas colors of Type 2s are a function of redshift.
At lower redshifts, Type 2s deviate farther from the single starburst track.
Fig. \ref{twocolor} shows that
Type 1 and Type 2 LAEs are discriminated more clearly in the diagram 
of z'-H vs J-K, compared to that of z'-K vs J-K.

\subsection{Brightness distribution of a Type 2 LAE}

\begin{figure*}
\includegraphics[width = 120mm]{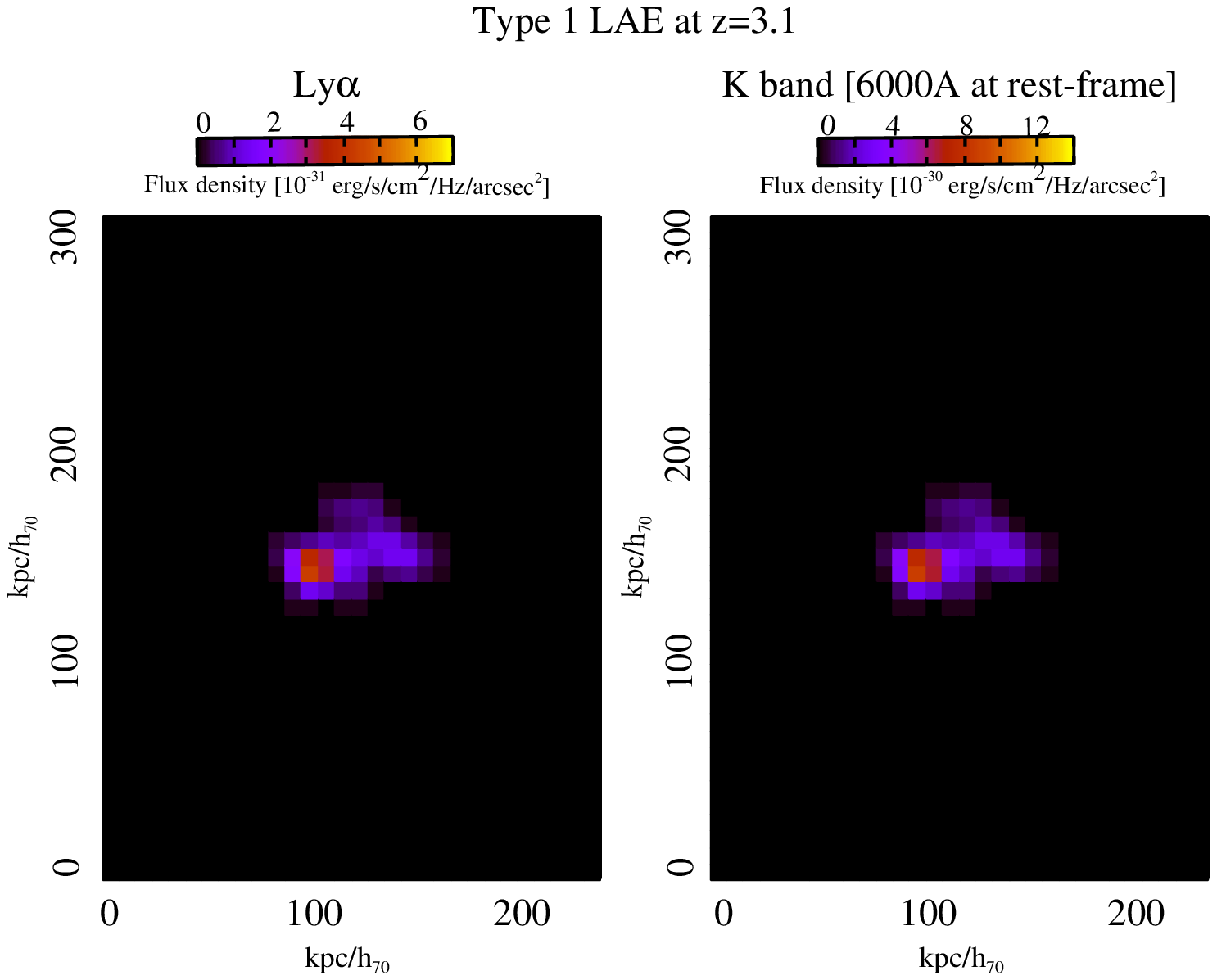}
\caption{The brightness distributions of a Type 1 LAE at $z=3.1$. 
Left and right panels show the brightness distributions of Ly$\alpha$ emission, 
and those at observed-frame K band flux (which corresponds to rest-frame 6000\AA\ flux), 
of stellar components, respectively. 
Each color bar shows the flux of Ly$\alpha$ emission and observed-frame K band flux, 
respectively. 
The angular resolution (pixel size) is set to be 1 arcsec.}
\label{brightdis1}
\end{figure*}

\begin{figure*}
\includegraphics[width = 120mm]{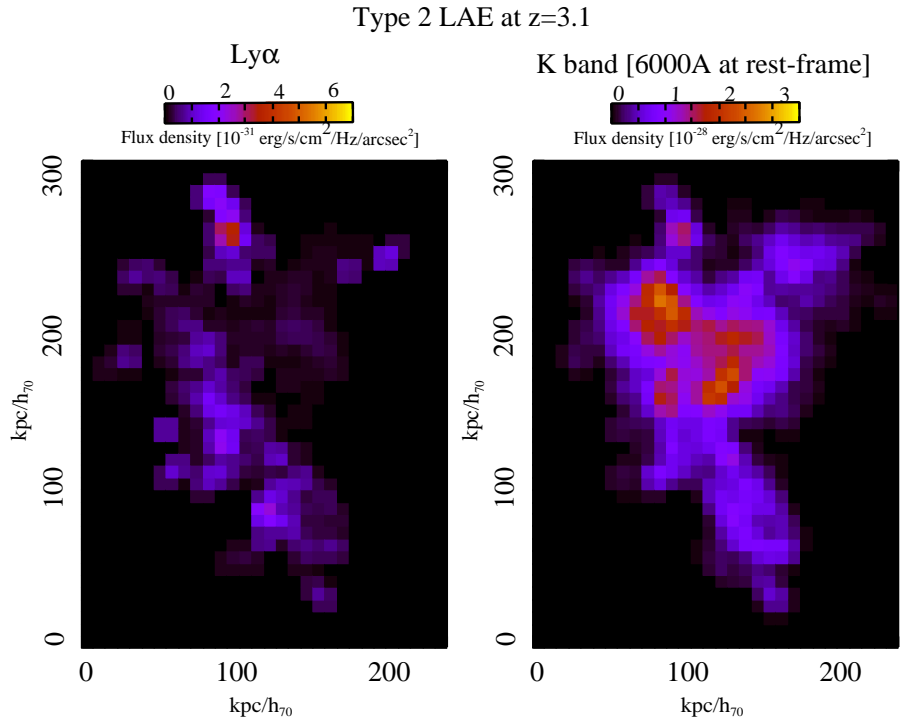}
\caption{Same as Fig. \ref{brightdis1}, but for a Type 2 LAE at $z=3.1$.}
\label{brightdis2}
\end{figure*}

\begin{figure}
\includegraphics[width = 80mm]{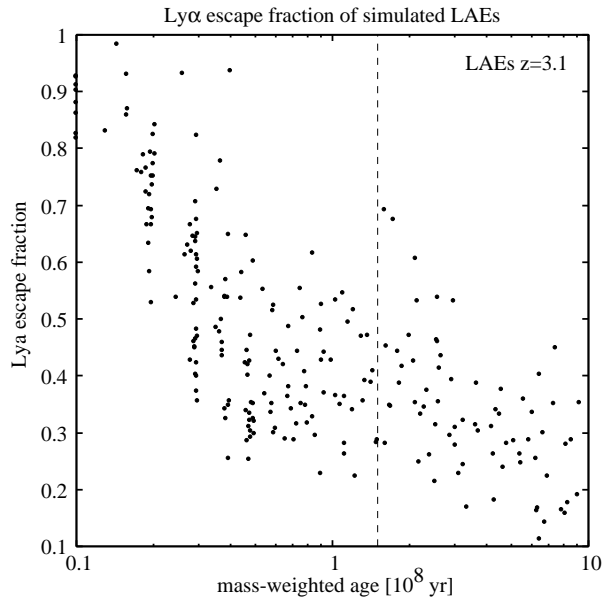}
\caption{Escape fractions of Ly$\alpha$ photons as a function of 
mass-weighted ages of LAEs at $z=3.1$. }
\label{escape}
\end{figure}

In Fig. \ref{brightdis1}, the brightness distributions of Ly$\alpha$ emission and 
those at observed-frame K band flux (which corresponds to rest-frame 6000\AA\ flux) 
are shown for a Type 1 LAE at $z=3.1$. 
The angular resolution (pixel size) is set to be 1 arcsec.
The brightness distributions at observed-frame K band flux trace basically stellar mass distribution. 
As seen in Fig. \ref{brightdis1}, both distributions are compact ($\approx 10h^{-1}$kpc), 
and the extents are quite similar to each other.
In Fig. \ref{brightdis2}, the brightness distributions are shown for a Type 2 LAE at $z=3.1$.
The Ly$\alpha$ emission is diffusely distributed over $\approx 100h^{-1}$kpc
or stronger at the outskirts,
while the brightness distributions at observed-frame K band flux 
exhibit a strong contrast and are more concentrated. 
Moreover, Type 2 LAE is composed of some clumps. 
Such clumpy structures can be seen some observed LABs in the SSA22 region 
\citep{Uchimoto2008, Webb2009}.
This result suggests that some Type 2 LAEs may not be well dynamically relaxed
after mergers, since the duration of strong Ly$\alpha$ emission is shorter than 
the relaxation time of such a large system.

Since Ly$\alpha$ emission is radiated mostly by starbursts, 
Ly$\alpha$ bright regions correspond to sites of subhalo accretion.  
These results suggest that if we observe LAEs in various wavelengths, 
the extent and morphology of Type 2 LAEs appear to be different in each band. 
Recently, such an offset has been reported in local starburst galaxy 
\citep{Hayes2007, Ostlin2009}, while no offset is found 
in LAEs at a higher redshift of $z = 4.4$ \citep{Fin2008}.
This can be related to the result that 
the number of more extended, older Type 2 LAEs 
increases at lower redshifts, as shown below.

Ly$\alpha$ emissions are affected more severely 
by dust extinction than continuum radiation. 
The effect of dust extinction is expected to be stronger for old metal-enriched systems.
In Fig. \ref{escape}, the escape fractions of Ly$\alpha$ photons are
plotted against ages of LAEs. The escape fractions are a decreasing function
of LAE ages as anticipated.
In particular, in Type 2 LAEs, the escape fractions decrease down to $\approx$ 10 percent.
Thus, for aged Type 2 LAEs, Ly$\alpha$ equivalent widths or luminosities  
become smaller than expected purely from star formation rate (see Fig. \ref{properties}).
Nevertheless, in the outskirts of halo, starbursts by delayed subhalo accretion 
take place frequently, where the dust extinction is relatively small.
Hence, Ly$\alpha$ emissions can be brighter in the outskirts of halo.

As shown in Fig. \ref{brightdis2},
the size of Type 2 LAEs can exceed $100~ \rm kpc$ in physical size. 
This is comparable to the size of Ly$\alpha$ blobs (LABs) \citep{Matsuda04}. 
Thus, some of Type 2 LAEs with EW$>$100\AA\, 
may account for a part of observed LABs.

\subsection{The halo-size distributions of simulated LAEs}

\begin{figure}
\includegraphics[width = 80mm]{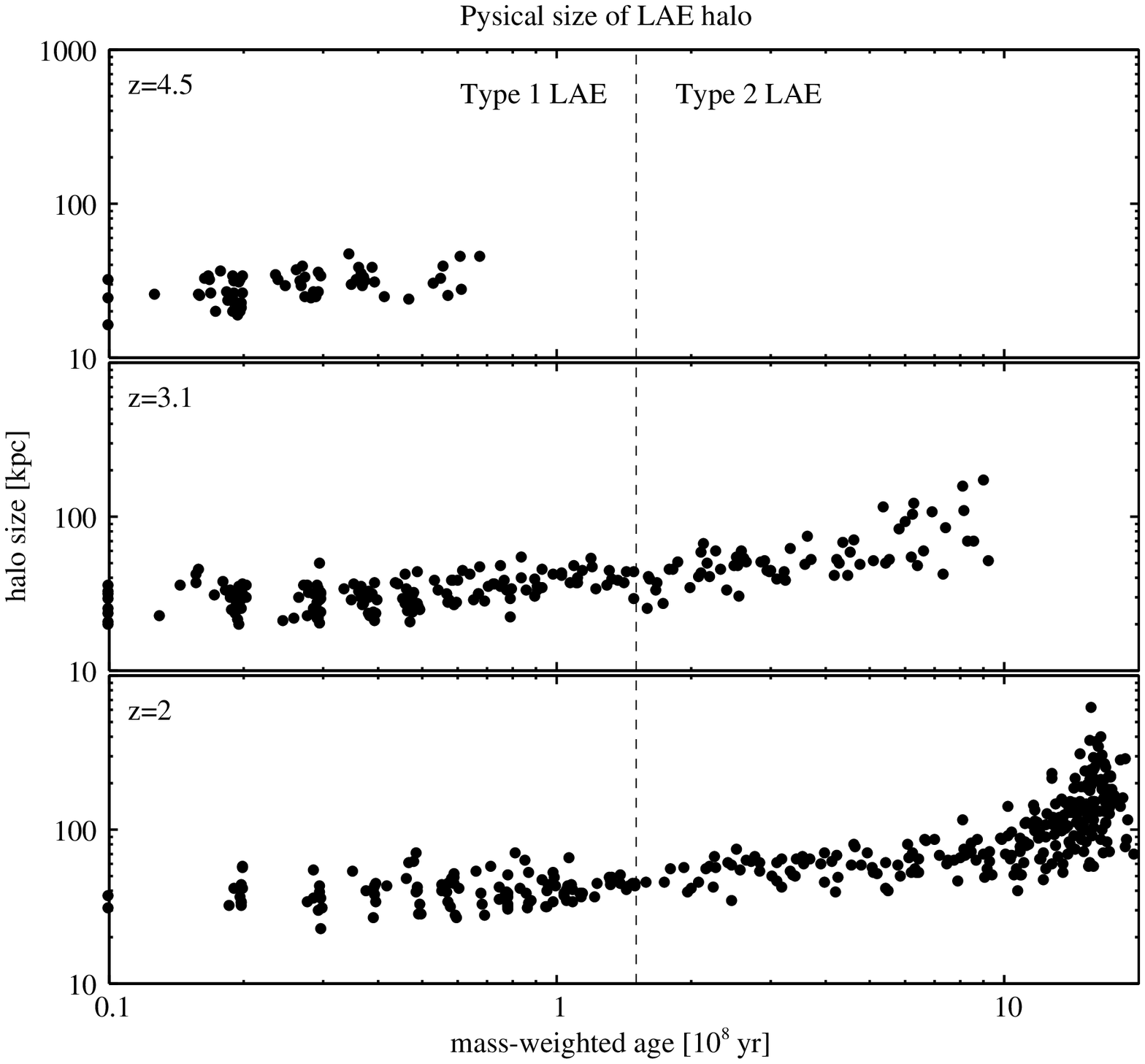}
\caption{The halo-size distributions of simulated LAEs 
as a function of mass weighted age at $z = 2, 3.1, $ and 4.5. }
\label{size}
\end{figure}

Here, we analyze the halo-size distributions of simulated LAEs. 
Fig. \ref{size} represents the halo-size distributions of all simulated LAEs 
as a function of mass wighted age at $z = 2, 3.1$, and 4.5. 
Here, the radius from the center of gravity of a halo within which $95 \%$ 
of the total mass is included is defined as the size of a halo. 
At $z = 4.5$, there are no Type 2 LAEs.
At lower redshifts, Type 2 LAEs  appear and
the number of more extended, older Type 2 LAEs  
increases with decreasing redshifts.
Interestingly, the range of Type 2 LAE size is quite broad.
Small Type 2 LAEs with the size of a few tens kpc are as compact as
Type 1 LAEs, while large Type 2 LAEs with the size of a few hundreds kpc
are comparable to Ly$\alpha$ blobs (LABs) \citep{Matsuda04}. 

\subsection{Clustering Properties of Two-type LAEs}

\begin{figure}
\includegraphics[width = 80mm]{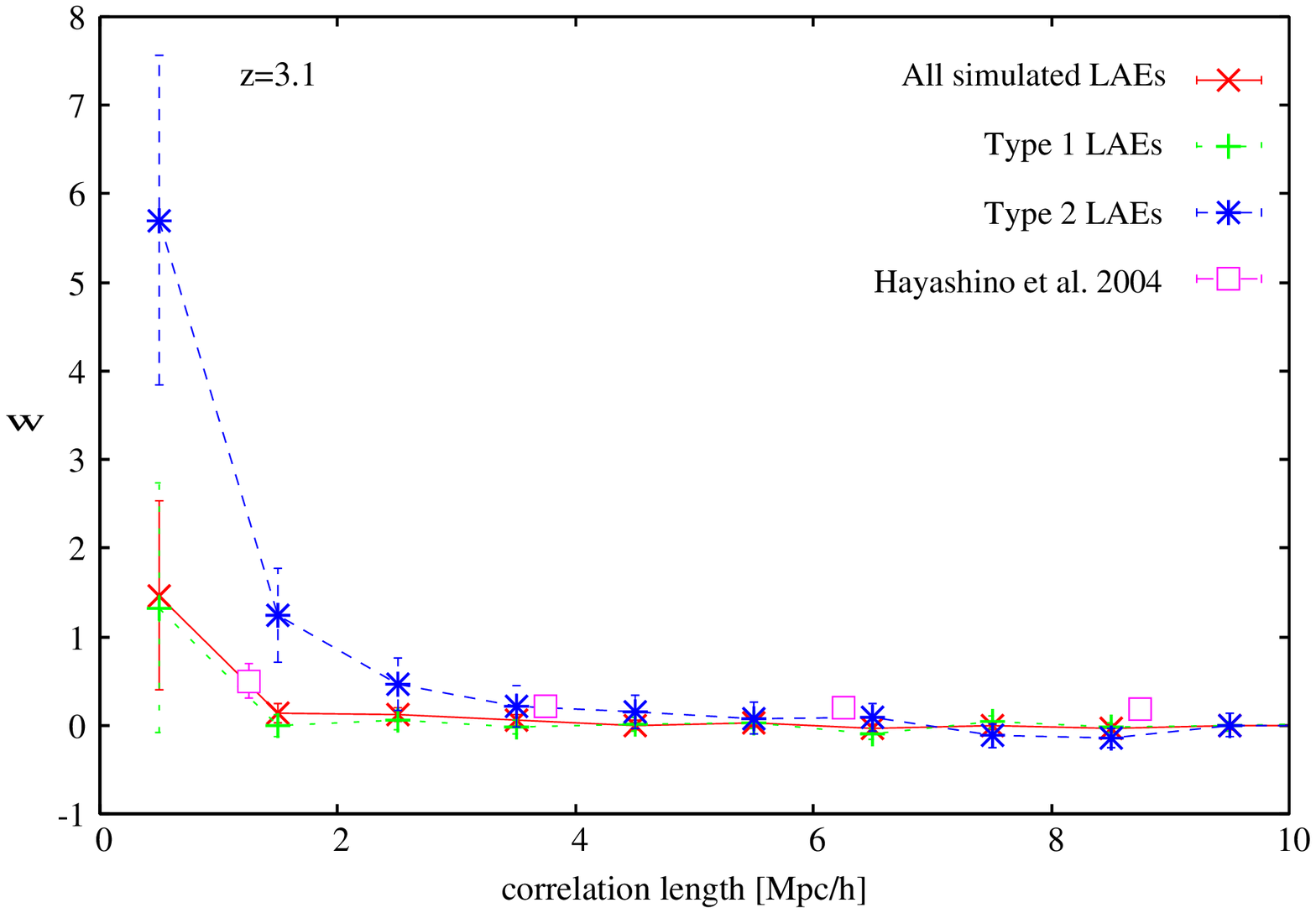}
\caption{Two-point angular correlation function (ACF) of all simulated LAEs 
and each type LAEs at $z=3.1$. 
A solid line is ACF of all simulated LAEs, 
and dashed and doted lines show ACF of Type 1 and Type 2 LAEs, respectively. 
Open-squares are the ACF of LAEs observed in SSA22a \citep{Ha2004}.}
\label{ACF}
\end{figure}

\begin{figure}
\includegraphics[width = 80mm]{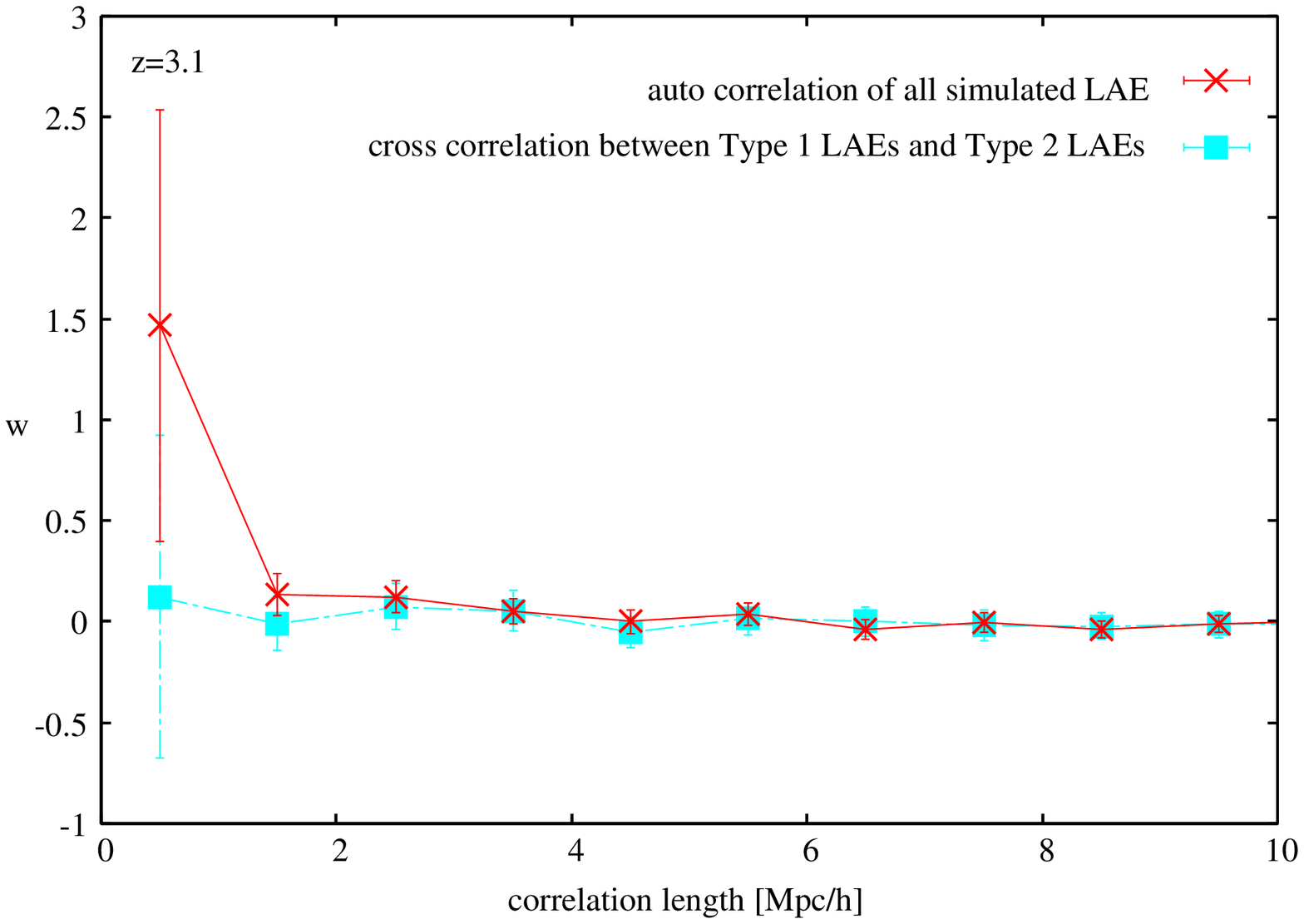}
\caption{Two-point angular cross-correlation function (CCF) of simulated LAEs. 
CCF between Type 1 and Type 2 LAEs is depicted by a dash-dot line.
For comparison, ACF of all simulated LAEs is also shown by a solid line.}
\label{CCF}
\end{figure}

In order to explore the clustering properties of two types of LAEs,
we calculate a two-point angular correlation function (ACF) and
a two-point angular cross-correlation function (CCF) between Type 1 and Type 2 LAEs.
Fig. \ref{ACF} represents ACFs of each type LAEs and all simulated LAEs. 
Also, the ACF of LAEs observed in the SSA22 field \citep{Ha2004} is shown. 
ACF of Type 1 LAEs as well as ACF of all simulated LAEs is quite weak
and well matches the observed ACF in in the SSA22 field,
whereas ACF of Type 2 LAEs is significantly stronger than that of Type 1 LAEs. 
This implies that Type 1 LAEs preferentially reside lower-density regions at $z=3.1$,
as shown by \citet{S2007}, while 
Type 2 LAEs are located in higher-density regions. 
The correlation strength of Type 2 LAEs is comparable to 
that of LBGs in SSA22 region \citep{Gia1998}. 
This result suggests that Type 2 LAEs may be a subsample of LBGs 
with strong Ly$\alpha$ emission. 

Fig. \ref{CCF} shows CCF between Type 1 and Type 2 LAEs.
The CCF is still weaker than ACF of all simulated LAEs.
Very recently, \citet{SMGLAE} have found that 
the cross correlation between Sub-mm galaxies (SMGs) and LAEs in the SSA22 field
is very weak. Type 2 LAEs possess a large amount of dust.
If many Type 2 LAEs can be detected as SMGs,
our simulation is consistent with this observation.

\section{Conclusions and Discussion}

To explore the origin of two populations of LAEs recently found,
we have performed three-dimensional cosmological $N$-body simulations 
of subhalo merging history in a $\Lambda$CDM universe. 
We have incorporated star formation history, SED evolution,
and dust extinction. 
As a result, we have found that the age of simulated LAEs can spread 
over a wide range from $2\times 10^6$yr to $9\times 10^8$yr. 
Also, we have revealed that there are two types of LAEs.
We have defined LAEs younger than $1.5 \times 10^8$yr as Type 1s,
and older ones as Type 2s.
In Type 1 LAEs early coeval starbursts occur in small parent halos,
while in Type 2 LAEs delayed starbursts take place in evolved massive haloes
as a consequence of delayed accretion of subhalos. 
A parent halo can experience repeatedly a Type 2 LAE-phase after 
a Type 1 LAE-phase.

The stellar mass of Type 1 LAEs is $M_{\rm star} \la 5 \times 10^8 \rm ~M_{\odot}$,
while Type 2 LAEs are as massive as 
$M_{\rm star} \sim 5 \times 10^8 - 3\times 10^{10} \rm ~M_{\odot}$.
The physical properties of Type 1 and Type 2 LAEs are
concordant with those of two populations of LAEs observed with
multi-wavelengths
\citep{Fin2007, Lai2008, Matsuda2007, Uchimoto2008, Fin2009}.
The fraction of Type 2s in all LAEs is a function of redshift,
which is less than 2 percent at $z \ga 4.5$,
$\sim$30 percent at redshift $z=3.1$, and 
$\sim$70 percent at $z=2$.
This trend is consistent with two populations of LAEs found by \citet{Nilsson2009}.  
Type 2 LAEs are expected to be discriminated clearly from Type 1 LAEs
in two color diagram of z'-H vs J-K.
We find that the brightness distribution of Ly$\alpha$ in Type 2 LAEs is 
more extended than the main stellar component, in contrast to Type 1 LAEs.
This is not only because delayed starbursts tend to occur in the outskirts of
a parent galaxy, but also because Ly$\alpha$ photons are effectively absorbed 
by dust in an evolved galaxy. 
The sizes of Type 2 LAEs range from a few tens to a few hundreds kpc.
At lower redshifts, the number of more extended, older Type 2 LAEs increases. 
Small Type 2 LAEs are as compact as Type 1 LAEs, while
large Type 2 LAEs exceeding $100~ \rm kpc$  
are comparable to Ly$\alpha$ blobs (LABs) \citep{Matsuda04}. 

Moreover, we have found that the clustering of Type 2 LAEs are even 
stronger than Type 1 LAEs. 
The amplitude of angular correlation function of Type 2 LAEs 
is comparable to that of Lyman break galaxies (LBGs) \citep{Gia1998}. 
This suggests that LBGs with strong Ly$\alpha$ line can be Type 2 LAEs.
The two-point angular cross-correlation function is still weaker 
than that of all LAEs. If many Type 2 LAEs can be detected as SMGs,
this result is consistent with recent observation by \citet{SMGLAE}.

Interestingly, in a low redshift universe at $0.2<z<0.35$, 
the {\it Galaxy Evolution Explorer (GALEX)} have found
many LAEs older than $2 \times 10^8$yr, more massive than
$10^9 \rm ~M_{\odot}$ in stellar component, and having 
small escape fractions of Ly$\alpha$ photons \citep{Deh08, Fin2009c}.
These properties are quite similar to those of Type 2 LAEs we
defined in this paper. 
Hence, local LAEs could be a good sample to study the detailed
physical states of Type 2 LAEs. 

Furthermore, in an evolved population of LAEs, 
active galactic nucleus (AGN) events may be anticipated,
since supermassive black holes reside evolved galaxies in a local universe
\citep[e.g.,][and references therein]{MH03}.
Recently, \citet{Yamada2009} have found that 
AGNs are exclusively $(90\%)$ associated with the massive objects 
with stellar mass larger than $10^{10.5}$ $\rm M_{\odot}$. 
Also, {\it GALEX} found that in low-redshift LAEs 
an AGN fraction is as high as forty percent \citep{Fin2009b}.
These findings may imply that Type 2 LAEs could have a higher fraction of AGNs.
Hence, it seems worth investigating the contribution of AGNs
to Ly$\alpha$ emission especially for Type 2 LAEs 
(Shimizu \& Umemura, in preparation).

\section*{Acknowledgments}
We are grateful to T. Hayashino, T. Yamada, Y. Matsuda and R. Yamauchi for providing 
valuable information and helpful comments, and 
to M. Mori and R. M. Rich for valuable discussion.
Numerical simulations have been performed with computational facilities 
at Center for Computational Sciences in University of Tsukuba,
and the EUP cluster system installed at Graduate School of Frontier Sciences, 
University of Tokyo. 
This work was partially supported by the {\it FIRST} project based on
Grants-in-Aid for Specially Promoted Research (16002003) and 
Grant-in-Aid for Scientific Research (S) (20224002),
and also by Grant-in-Aid for Young Scientists (S) (20674003).

\bsp

\label{lastpage}

\end{document}